\documentclass{aa}

\usepackage{setspace}
\usepackage{graphicx}
\usepackage{float}
\usepackage{amsmath}
\usepackage{threeparttable}
\usepackage{txfonts}
\usepackage{natbib}
\usepackage{subfigure}
\usepackage{gensymb}
\usepackage{bm}
\usepackage[justification=centering]{caption}

\usepackage{overpic}

\begin{document} 

\title{Distribution of dust ejected from the lunar surface into the Earth-Moon system}
   \author{Kun Yang\inst{1,3}
          \and Jürgen Schmidt\inst{2,3}
          \and Weiming Feng\inst{1}
          \and Xiaodong Liu\inst{4}
          }
   \institute{Department of Engineering Mechanics, Shandong University,
              250061 Jinan, China
         \and
            Institute of Geological Sciences, Freie Universität Berlin, Berlin, Germany
         \and
             Space Physics and Astronomy Research Unit, University of Oulu, 90014 Oulu, Finland
         \and
             School of Aeronautics and Astronautics, Sun Yat-sen University, Shenzhen Campus, 518107 Shenzhen, China \\
             \email{liuxd36@mail.sysu.edu.cn}
             }
   \abstract
   {}
   {An asymmetric dust cloud was detected around the Moon by the Lunar Dust Experiment on board the Lunar Atmosphere and Dust Environment Explorer mission. We investigate the dynamics of the grains that escape the Moon and their configuration in the Earth-Moon system.}
   {We use a plausible initial ejecta distribution and mass production rate for the ejected dust. Various forces, including the solar radiation pressure and the gravity of the Moon, Earth, and Sun, are considered in the dynamical model, and direct numerical integrations of trajectories of dust particles are performed. The final states, the average life spans, and the fraction of retrograde grains as functions of particle size are computed. The number density distribution in the Earth-Moon system is obtained through long-term simulations.}
   {The average life spans depend on the size of dust particles and show a rapid increase in the size range between $1\, \mathrm{\mu m}$ and $10\, \mathrm{\mu m}$. About ${3.6\times10^{-3}\,\mathrm{kg/s}}$ ($\sim2\%$) particles ejected from the lunar surface escape the gravity of the Moon, and they form an asymmetric torus between the Earth and the Moon in the range $[10\,R_\mathrm{E},50\,R_\mathrm{E}]$, which is offset toward the direction of the Sun. A considerable number of retrograde particles occur in the Earth-Moon system.}
   {}
              
   \titlerunning{Dust particles from the lunar surface}
   \authorrunning{Yang et al.}         
   \maketitle
\section{Introduction}

   In the past, studies on the configuration of circumplanetary dust mainly focused on the grain particles around the giant planets and Mars \citep[e.g.,][]{tiscareno2018planetary, Spahn2019Circumplanetary}. The dynamics of dust particles in the Earth-Moon system has been less explored. Although the Munich Dust Counter did not detect the lunar ejecta dust cloud \citep{iglseder1996cosmic}, a permanent and asymmetric dust cloud engulfing the Moon was indeed detected by the Lunar Dust Experiment \citep[LDEX;][]{horanyi2015permanent}. From the LDEX measurements, the size distribution of the dust grains was derived, and the density of particles was found to drop with altitude and to vary azimuthally around the Moon, reaching a peak at 5-7 hours lunar local time. Six previously identified meteoroid populations, helion, antihelion, apex, antiapex, northern toroidal, and southern toroidal, were found to be plausible interplanetary projectiles that generate the lunar dust cloud via high-speed impacts \citep{szalay2019impact}. The asymmetric nature of this dust cloud, the result of the asymmetric impactor flux, was analyzed, and a consistent mass production rate per unit surface as a function of impact direction was derived. \citet{Szalay2016Lunar} also analyzed particles that fall back to the lunar surface, which follow a cumulative size distribution with exponent 2.7. The dynamics of dust particles ejected from the Moon was previously studied by \citet{colombo1966earth}, and dust orbiting the near-Earth environment was studied by \citet{Peale1966Dust}. 
   
   Investigating the lunar dust distributions at different times, heights, and positions was ever one of optional tasks of Chinese Lunar Exploration Mission, Chang'e 4 \citep{Wang2016A}. At the end of 2020, Chang'e 5 launched successfully. The dust distribution can be obtained from dust dynamics modeling; as such, the total dust flux received by the spacecraft can be calculated, and the dust hazard for a specific mission can be evaluated. Space activities in the Earth-Moon system are much more frequent compared with other planets, and thus the study of dust particles is of great significance for assessing the space environment and ensuring the security of explorations.
    
   In this work, we focus on the dynamics and distribution of particles ejected from the surface of the Moon that escape the Moon's gravity. The paper is structured as follows. In Sect. 2 the dynamical model for the motion of particles ejected from the lunar surface is presented, taking the solar radiation pressure and gravity of the Sun, the Earth, and the Moon into consideration. In Sect. 3 the mass production rate and the initial ejecta distribution on the surface of the Moon are deduced. In Sect. 4 the detailed simulation results are presented, with the distribution of sinks and the average life spans of particles as functions of size. The dust number density distribution in the Earth-centered inertial frame is also presented. Finally, the distributions and the evolution of osculating orbital elements of dust particles are given.   
\section{Dynamical model}

 Dust particles in interplanetary and circumplanetary space can be affected by various forces, including solar gravity, solar radiation pressure, Poynting-Robertson drag, the Lorentz force, and the gravity of the planets and other bodies in the Solar System. For particles ejected from the surface of the Moon, solar radiation pressure and the gravity of the Moon, Earth, and Sun are considered in this paper. We estimated the strength of the perturbation induced by the Lorentz force for typical strengths of the magnetic field and grain charges in the interplanetary magnetic field, the Earth’s plasma sphere, and the geomagnetic tail of the Earth. We find that outside $10\,R_\mathrm{E}$ the strength of the Lorentz force is small compared to the perturbations induced by solar radiation pressure, by at least two orders of magnitude for grains larger than $10 \, \mathrm{\mu m}$; even for grains of $1\,\mathrm{\mu m}$, solar radiation pressure dominates by more than an order of magnitude. An Earth-centered $\mathrm{J}2000$ inertial frame $Oxyz$ is used in our simulation. Here, the $x$ axis points in the direction of vernal equinox at the $\mathrm{J}2000$ epoch, the $z$ axis is the normal of the Earth's equatorial plane (north), and the $y$ axis is determined by the right-handed rule. 
 
\begin{figure}
    \centering
    \includegraphics[width=7.5cm,height=5.5cm]{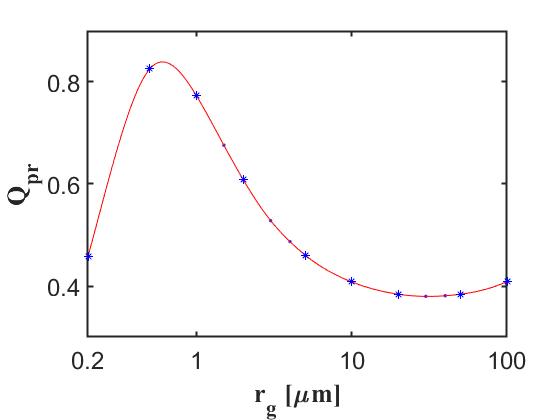}
    \caption{Solar radiation pressure efficiency, $Q_{\mathrm{pr}}$, for spherical silicate particles calculated from Mie theory.}
    \label{fig: Qpr}
\end{figure}

 Physical parameters for the Sun, Earth, and Moon used in this paper are shown in Table \ref{table:1}. All the ephemerides of the Sun, Earth, and Moon in our model are taken from NAIF SPICE toolkit\footnote{http://naif.jpl.nasa.gov}.
\begin{table}[H]
\centering  
\caption{Physical and orbital properties of celestial bodies.}             
\label{table:1}      
\begin{tabular}{c c c c }        
\hline\hline                 
 & $M\left(\mathrm{kg}\right)$ & $R\left(\mathrm{km}\right)$   & $v_{\mathrm{esc}}\left(\mathrm{km}/\mathrm{s}\right)$ \\

 \hline                        
            Sun &$1.989\times10^{30}$  &$6.955\times10^5$   &617.70  \\
            Earth &$5.972\times10^{24}$ &$6.378\times10^3$  &11.180 \\
            Moon &$7.347\times10^{22}$ &$1.737\times10^3$ &2.3416\\   
\hline                                   
\end{tabular}
 \begin{tablenotes}
 \centering
        \footnotesize
        \item[a] $M$, $R$, and $v_{\mathrm{esc}}$ denote mass, radius, and escape velocity. 
      \end{tablenotes}
\end{table}
The equation of motion of one particle ejected from the surface of the Moon reads
\begin{equation}
\ddot{\vec{r}} = \ddot{\vec{r}}_{\mathrm{G}_{\mathrm{E}}}+\ddot{\vec{r}}_{\mathrm{RP}}+\ddot{\vec{r}}_{\mathrm{G}_{\mathrm{others}} }
,\end{equation}
where $\vec{r}$ is the Earth-centric radius vector of the dust particle and $\ddot{\vec{r}}_{\mathrm{G}_{\mathrm{E}}}$ is the acceleration caused by the gravity of the Earth,
\begin{equation}
\ddot{\vec{r}}_{\mathrm{G}_{\mathrm{E}}}=GM_E\nabla{\left\{\frac{1}{r}\left[1-{J_2}\left(\frac{R_\mathrm{E}}{r}\right)^2P_2(\cos\theta)\right]\right\}}
.\end{equation}
Here, $G$ is the gravitational constant, $M_\mathrm{E}$ is the mass of the Earth, $R_\mathrm{E}$ is the reference radius of Earth, $\theta$ is the colatitude in an Earth-centered body-fixed frame, and $P_2$ is the Legendre function of degree $2$. In our simulation only the second-degree zonal harmonic, $J_2 \approx 1.082 \times 10^{-3} $ \citep{pavlis2012development}, is considered. The variable $\ddot{\vec{r}}_{\mathrm{RP}}$ denotes the acceleration due to the solar radiation pressure \citep{burns1979radiation},
\begin{equation}
    \ddot{\vec{r}}_{\mathrm{RP}}=\frac{3Q_\mathrm{c}Q_{\mathrm{pr}}\mathrm{AU}^2}{4\left(\vec{r}-\vec{r_\mathrm{S}}\right)^2\rho_\mathrm{g}r_\mathrm{g}c} \left[1-\frac{\left(\dot{\vec{r}}-\dot{\vec{r_\mathrm{S}}}\right)\cdot{\vec{\hat{r}_{\mathrm{Sp}}}}}{c}\right]\vec{\hat{r}_{\mathrm{Sp}}}
,\end{equation}
where $Q_\mathrm{c}=1.36 \times10^3 \, \mathrm{W}/\mathrm{m^2}$ is the solar radiation energy flux at one astronomical unit ($\mathrm{AU}$) distance and $Q_{\mathrm{pr}}$ is the solar radiation pressure efficiency factor, which depends on the size and material of the dust particle. The value of $Q_{\mathrm{pr}}$ was calculated from Mie theory for spherical particles using optical constants for silicates taken from \citet{Mukai1989} (see our Fig.~\ref{fig: Qpr}). The symbol $\rho_\mathrm{g}=3500\, \mathrm{kg/m^3}$ denotes volumetric mass density for lunar dust \citep{solomon1974density}, $r_\mathrm{g}$ is the radius of the particle, $c$ is the speed of light, $\vec{r_\mathrm{S}} $ is the vector from the Earth to the Sun, and $\hat{\vec{r_{\mathrm{Sp}}}}$ is the unit vector from the Sun to the particle.

The acceleration by the gravity of Sun and Moon reads \citep[e.g.,][]{murray1999solar, liu2019dust}
\begin{equation}
\ddot{\vec{r}}_{\mathrm{G}_{\mathrm{others}}}=GM_\mathrm{S}\left(\frac{\vec{r_{\mathrm{pS}}}}{r_{\mathrm{pS}}^3}-\frac{\vec{r_{\mathrm{S}}}}{r_{\mathrm{S}}^3}\right)+GM_\mathrm{M}\left(\frac{\vec{r_{\mathrm{pM}}}}{r_{\mathrm{pM}}^3}-\frac{\vec{r_{\mathrm{M}}}}{r_{\mathrm{M}}^3}\right) 
,\end{equation}
where $M_\mathrm{S}$ and $M_\mathrm{M}$ are the masses of the Sun and the Moon. The symbol $\vec{r_{\mathrm{pS}}} $ denotes the vector from the dust particle to the Sun, $\vec{r_{\mathrm{pM}}} $ is the vector from the dust particle to the Moon, and $\vec{r_{\mathrm{M}}} $ is the vector from the Earth to the Moon.

In Sect. 4 we analyze the contribution of different perturbation forces to the evolution of eccentricity and inclination, $\mathrm{d}e/\mathrm{d}t$ and $\mathrm{d}i/\mathrm{d}t$, which can be calculated from the perturbation equations as \citep{murray1999solar}
\begin{equation}
\frac{\mathrm{d}e}{\mathrm{d}t}=\sqrt{a\mu^{-1}\left(1-e^2\right)}\left[\overline{R}\sin f + \overline{T}\left(\cos f + \cos E\right)\right]
\end{equation}
\begin{equation}
\frac{\mathrm{d}i}{\mathrm{d}t}=\frac{r\overline{N}\,\cos\left(\omega+f\right)}{h}
.\end{equation}
Here, $\overline{R}$, $\overline{T}$, and $\overline{N}$ are the magnitudes of the radial, tangential, and normal components of the acceleration, respectively, which we determined from the equations of motion (Eq. 1). The symbols $a, e, i, \omega, f$, and $ E$ denote the osculating semimajor axis, eccentricity, inclination, argument of periapsis, true anomaly, and eccentric anomaly, respectively. We used the standard expressions from the two-body problem for radial distance,
\begin{equation}
r=\frac{a\left(1-e^2\right)}{1+e\cos f}
,\end{equation}
and angular momentum per unit mass,
\begin{equation}
h=\sqrt{\mu a \left(1-e^2\right)}
.\end{equation}
The integration of a trajectory is terminated in our simulation if the grain hits the surface of the Earth or the Moon or if it escapes from the Earth-Moon system. We did not consider the effect of the Earth's atmosphere. Collisions with the Earth or Moon could be easily missed due to the discrete time steps of the integrator. Therefore, cubic Hermite interpolation is used in our model to ensure the detection of collisions with the Earth or Moon if they occur between two consecutive time steps of the integrator \citep{liu2016dynamics}. 

\section{Mass production rate and initial ejecta mass distribution}

For the starting velocity we adopted a simple model in which the ejection direction is perpendicular to the lunar surface and all the particles have the same velocity distribution regardless of size. We assumed that the starting velocity follows a power law with exponent $q$. By normalizing $\int_{v_0}^{v_{\mathrm{max}}}{p(v)}\mathrm{d}v$, we obtain the form of the initial velocity distribution,
\begin{equation}
p(v)=\frac{1-q}{v_{\mathrm{max}}^{1-q}-v_0^{1-q}} {v}^{-q}
,\end{equation}
where $q=2.2$ was used, suitable for a regolith-covered surface \citep{krivov2003impact}, $v_0=130\, \mathrm{m/s}$ is the minimum speed for lunar ejecta \citep{horanyi2015permanent}, and $v_{\mathrm{max}}=2v_\mathrm{esc}\approx4.68 \, \mathrm{km/s}$ is adopted for the maximum ejection speed.

Via normalization to the total mass production rate, $\int_{m_{\mathrm{min}}}^{m_{\mathrm{max}}}{mp\left(m\right)\mathrm{d}m}=M_\mathrm{total}^+$, the distribution of mass can be expressed as
\begin{equation}
p(m)=M_\mathrm{total}^+\frac{1-\alpha}{m_{\mathrm{max}}^{1-\alpha}-m_{\mathrm{min}}^{1-\alpha}}m^{-\left(1+\alpha\right)}
,\end{equation}
where $\alpha=0.9$ was inferred from the LDEX data \citep{horanyi2015permanent}, $m_{\mathrm{min}}$ is the minimum ejecta mass, and $m_{\mathrm{max}}$ is the maximum ejecta mass. The ejecta size-distribution is obtained from the transformation $p\left(m\right)\mathrm{d}m=p(r_\mathrm{g})\mathrm{d}r_\mathrm{g}$ \citep[e.g.,][]{liu2018dust},
\begin{equation}
p(r_\mathrm{g})=\frac{3}{r_{\mathrm{max}}}\frac{M_\mathrm{total}^+}{m_{\mathrm{max}}}\frac{1-\alpha}{1-\left(\frac{m_{\mathrm{min}}}{m_{\mathrm{max}}}\right)^{1-\alpha}}\left(\frac{r_\mathrm{g}}{r_{\mathrm{max}}}\right)^{-1-3\alpha}
,\end{equation}
where $r_{\mathrm{max}}=\sqrt[3]{\frac{3m_\mathrm{max}}{4\pi\rho_\mathrm{g}}}$. 

To apply these distributions of mass or radius, we must specify the total mass production rate, $M_\mathrm{total}^+$. For the first step, the mass production rate per unit surface is 
\begin{equation}
\overline{M^+}=m_{\mathrm{p}}F_{\mathrm{p}}\cos{\phi}Y_\mathrm{p} 
,\end{equation}
where $F_\mathrm{p}$ is the number flux of impactors with characteristic mass $m_\mathrm{p}$, treated here as a parallel beam that hits the surface with an angle $\phi$ from the surface normal, and $\cos\phi$ is the projection area factor. The yield ,$Y_\mathrm{p}$, is defined as the ratio of the ejecta particles' mass to the impactors' mass, which is also a function of the material of the target surface as well as the mass and velocity of the impactors \citep{koschny2001impactsa}.

Considering the effects of oblique impacts from experiments by \citet{gault1973displaced} and the dependence on projectile mass and velocity, we have
\begin{equation}
Y_\mathrm{p}\propto m_\mathrm{p}^{0.23} v_\mathrm{p}^{2.46}\cos^2\phi
.\end{equation}
The mass production on the surface of the Moon varies with local time because the angle $\phi$ varies for projectile populations that approach the Earth-Moon system from different directions. Six meteoroid populations, helion (HE), antihelion (AH), apex (AP), antiapex (AA), northern toroidal (NT), and southern toroidal (ST), are found to be plausible sources of high-speed impacts \citep{szalay2019impact}. Following \citet{Szalay2015Annual}, and taking Eqs. (13) and (14) into account, the mass production rate per unit surface reads
\begin{equation}
\overline{M^\mathrm{+}}=C\sum_s\underbrace{F_{s}m_{s}^{1.23}v_{s}^{2.46}}_{w_s} \cos^3\phi_s\Theta_\mathrm{H}\left(\cos\phi_s\right) 
.\end{equation}
Here, $C$ is a normalization constant, $s$ labels the different projectile populations, $F_s$, $m_s$, and $v_s$ are the number flux, characteristic mass, and impact velocity for each source, and $\Theta_\mathrm{H}$ denotes the Heaviside function, which accounts for the fact that a given projectile population can only reach one hemisphere of the Moon. We followed \citet{szalay2019impact} and used relative contributions $w_s=(0.198,0.198,0.303,0.025,0.138,0.138)$ for the HE, AH, AP, AA, NT, and ST sources of meteoroids. The symbol $\phi_s$ is the impact angle for each source from the surface normal,
\begin{equation}
\cos\phi_s=\sin\theta\sin\theta_s\cos(\varphi-\varphi_s)+\cos\theta\cos\theta_s
.\end{equation}
Here, $\theta$ is the colatitude of the surface element, $\theta_s$ is the colatitude for sources, $\theta_{\mathrm{NT}}=30^\circ, \theta_{\mathrm{ST}}=150^\circ$ with $\varphi_{NT}=\varphi_{ST}=0^\circ$ \citep{szalay2019impact} , $\varphi$ is defined as the longitude from the apex of the motion of the Earth-Moon system for the surface element, $\varphi_s$ is the source-specific longitude of approach in the ecliptic plane (HE, AH, AP, and AA), $\varphi_{\mathrm{HE}}=65^\circ, \varphi_{\mathrm{AH}}=295^\circ,\varphi_{\mathrm{AP}}=0^\circ$, and $\varphi_{\mathrm{AA}}=180^\circ$ \citep{Szalay2016Lunar}. Therefore, Eq.~$(14)$ can be written as a function of the longitude, $\varphi$, and colatitude, $\theta$, on the Moon. The total mass production rate, $M_{\mathrm{total}}^+$, for the Moon was estimated to be approximately $0.2\, \mathrm{kg/s}$ \citep{szalay2019impact,pokorny2019meteoroids}. By normalization, $\int_{0}^{\pi}{R_M^2\sin\theta\mathrm{d}
\theta}\int_{0}^{2\pi}{\mathrm{d}\varphi \overline{M^+}(\varphi,\theta)}=M_\mathrm{total}^+$, the normalization constant, $C$, in Eq. $(14)$ can be fixed.

\begin{figure}
    \centering
    \includegraphics[width=8.5cm,height=6.0cm]{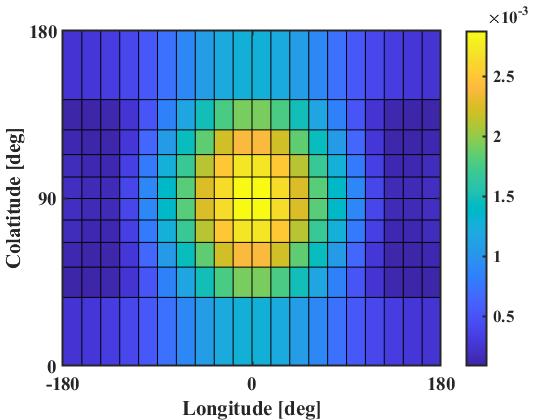}
    \caption{Mass production rate per surface element on the lunar surface. Zero longitude is defined by the direction of the apex of the motion of the Earth-Moon system.}
    \label{fig: MassProductionDistri}
\end{figure}

In our simulation, $200$ starting positions on the surface of the Moon, including $20$ longitudes and $10$ colatitudes, are selected, which divides the whole surface of the Moon into $20\times10$ elements. Longitudes, $\varphi$, are equidistant from $0$ to $2 \pi$, and colatitudes, $\theta$, vary from $0$ to $\pi$. Accordingly, the mass production rate for a given surface element $(i,j)$ reads
\begin{equation}
M^+(i,j)=\int_{\theta_j^1}^{\theta_j^2}{R_M^2\sin\theta}\mathrm{d}{\theta}\int_{\varphi_i^1}^{\varphi_i^2}{\mathrm{d}{\varphi}} \overline{M^+}(\varphi,\theta)
,\end{equation}
where $i$ is the longitude index from $1$ to $20$, $j$ is the colatitude index from $1$ to $10$, $R_M$ is the radius of the Moon, $\varphi_i^1$ and $\varphi_i^2$ are boundary longitudes of surface element $(i,j)$, and $\theta_j^1$ and $\theta_j^2$ are boundary colatitudes of surface element $(i,j)$. The distribution of mass production rates is shown in Fig.~\ref{fig: MassProductionDistri}.
Correspondingly, the size distribution for surface element $(i,j)$ reads
\begin{equation}
    p(r_\mathrm{g},i,j)=\frac{3}{r_{\mathrm{max}}}\frac{M_\mathrm{+}(i,j)}{m_{\mathrm{max}}}\frac{1-\alpha}{1-\left(\frac{m_{\mathrm{min}}}{m_{\mathrm{max}}}\right)^{1-\alpha}}\left(\frac{r_\mathrm{g}}{r_{\mathrm{max}}}\right)^{-1-3\alpha}
.\end{equation}

\section{Simulation results }
To cover the size distribution of ejecta, we selected nine dust particle radii: $0.2\, \mathrm{\mu m}$, $0.5\, \mathrm{\mu m}$, $1\, \mathrm{\mu m}$, $2\, \mathrm{\mu m}$, $5\, \mathrm{\mu m}$, $10\, \mathrm{\mu m}$, $20\, \mathrm{\mu m}$, $50\, \mathrm{\mu m}$, and $100\, \mathrm{\mu m}$. Here, $r_\mathrm{min}=0.2\,\mathrm{\mu m} $ and $\, r_\mathrm{max}=100\,\mathrm{\mu m}$, so $m_\mathrm{min}=\frac{4}{3}\rho_\mathrm{g}\pi r_\mathrm{min}^3 $ and$\,m_\mathrm{max}=\frac{4}{3}\rho_\mathrm{g}\pi r_\mathrm{max}^3$. For the initial velocity, ten velocities in the range $[0.95\,v_{\mathrm{esc}},2\,v_{\mathrm{esc}}]$ were used, where $v_{\mathrm{esc}}$ is the escape velocity of the Moon. We note that the velocity distribution is normalized to unity in the range $[v_0, v_\mathrm{max}=2 v_\mathrm{esc}]$, where $v_0=130 \, \mathrm{m/s}$ is the minimal starting velocity of ejecta. Integrating over the velocity distribution in the interval $[0.95v_\mathrm{esc},2 v_\mathrm{esc}]$ then removes in our normalization the grains in $[v_0,0.95v_\mathrm{esc}]$, which all fall back to the surface. A small fraction of grains in the range $[0.95v_\mathrm{esc},v_\mathrm{esc}]$ escape the Moon due to three-body effects. They are considered in our evaluation of the simulations. 

To average over seasonal effects, we assumed that particles start at $12$ different times, covering equidistantly one period of the ascending node precession of the Moon. We verified that this choice also covers the different phases of the Moon fairly uniformly. The number of grains necessary to cover the ranges of radius, velocity, time, colatitude, and longitude are $9$, $10$, $12$, $10$, and $20$, respectively. Thus, the total number of particles is $216,000$, which requires a huge amount of CPU time. The long-term simulations were carried out on the supercomputer located at the Finnish CSC-IT Center for Science. 

\subsection{Particle sinks and particle lifetimes}

The final state of particles ejected from the surface of the Moon after $100$ years is shown in Table \ref{table:2}. Most dust particles ($>88\%$) ultimately leave the Earth-Moon system after $100$ Earth years, regardless of size. 
\begin{table}[H]
\caption{Fate of particles ejected from the lunar surface after $100$ years (fraction of $24,000$ particles for each size).}             
\label{table:2}      
\centering  
\scalebox{0.90}{
\begin{tabular}{c c c c c c}  
\hline\hline                 
$r_g \, [\mathrm{\mu m}]$ &Hit Moon &Hit Earth & In Orbit & Escape \\
 \hline                        
        0.2 &0.00E+0 &1.18E-2 &0.00E+0 &9.88E-1\\
        0.5 &5.30E-4 &1.55E-2 &0.00E+0 &9.84E-1\\
        1   &4.15E-3 &4.30E-2 &0.00E+0 &9.53E-1\\
        2   &6.69E-3 &1.05E-1 &0.00E+0 &8.88E-1\\
        5   &1.47E-2 &8.91E-2 &5.34E-5 &8.96E-1\\
        10  &1.79E-2 &5.47E-2 &3.75E-4 &9.27E-1\\
        20  &2.22E-2 &4.00E-2 &5.90E-4 &9.37E-1\\
        50  &2.50E-2 &3.74E-2 &6.98E-4 &9.37E-1\\
        100 &2.63E-2 &3.56E-2 &6.44E-4 &9.38E-1\\

\hline                                   
\end{tabular}
}
\end{table}

Only a very small fraction of grains ($>2\, \mathrm{\mu m}$) remain in orbit after $100$ years. We do not integrate these trajectories further, to save CPU time. The fraction of grains that rapidly re-impact the Moon within one orbital period is approximately $22\%$, which only weakly depends on grain size. We note that this fraction depends on our choice of $v_{\mathrm{min}}=0.95v_{\mathrm{esc}}$, so the fraction of grains falling back to the Moon within one orbital period is not listed in Table \ref{table:2}. The grains that re-impact the Moon after a longer time are listed in Table \ref{table:2} in the column ``Hit Moon.'' Small particles are more sensitive to solar radiation pressure. The fraction of grains that hit the Moon increases with grain size. The fraction that hit the Earth grows from $1.2\%$ to $10\%$ at a grain size of $2 \, \mathrm{\mu m}$ and then begins to drop to a value of $3\%$ for $100 \, \mathrm{\mu m}$ grains. 

\begin{figure}
    \centering
    \subfigure[]
    {
    \includegraphics[width=7.5cm,height=5.5cm]{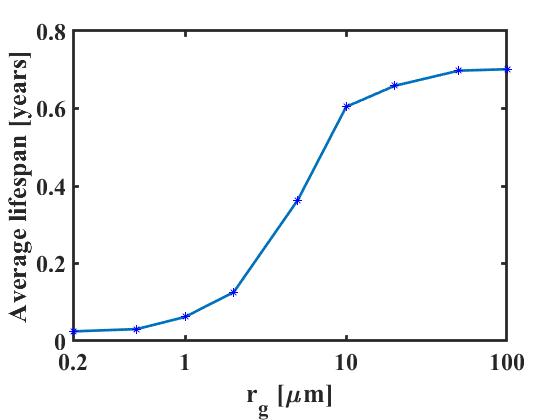}
    }
    \subfigure[]
    {
    \includegraphics[width=7.5cm,height=5.5cm]{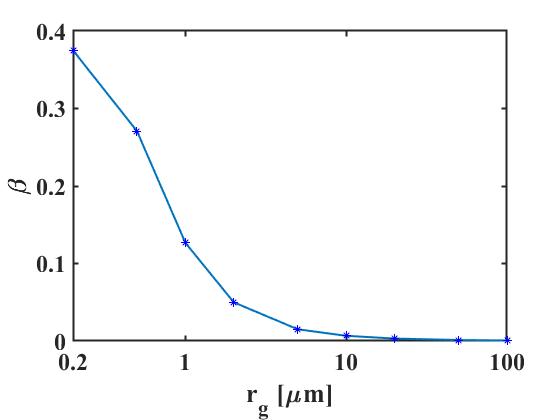}
   }
    \caption{Average lifetimes and $\beta$ for grains of different sizes. {\emph{Panel a}: Average lifetimes. \emph{Panel b}: $\beta$.}}
    \label{fig: LifeSpan}
\end{figure}

{Figure}~\ref{fig: LifeSpan}(a) shows the average life spans as a function of the size of the dust particles. Small particles, including those sized $0.2\, \mathrm{\mu m}$, $0.5\, \mathrm{\mu m}$, and $1 \, \mathrm{\mu m}$, have short lifetimes, less than $0.1$ years. For particles between $1 \, \mathrm{\mu m}$ and $10 \, \mathrm{\mu m}$, the life span increases rapidly with grain size. Large particles with sizes bigger than $10 \, \mathrm{\mu m}$ have longer lifetimes, approximately $0.7$ Earth years. The parameter $\beta$ denotes the ratio of solar radiation pressure to gravity \citep{burns1979radiation}:
\begin{equation}
\beta=\frac{3Q_\mathrm{c}Q_{\mathrm{pr}}\mathrm{AU}^2}{4GM_\mathrm{s}\rho_\mathrm{g}r_\mathrm{g}c} 
.\end{equation}

As shown in Fig.~\ref{fig: LifeSpan}(b), the effect of solar radiation pressure is large for small particles, which induces high eccentricities, such that small grains are rapidly expelled from the Earth-Moon system and the life spans of those particles are short. With increasing grain size, $\beta$ decreases rapidly and becomes negligible for grains of $10 \, \mathrm{\mu m}$ or larger. Thus, the lifetimes of these particles are much longer.
\subsection{Steady-state configuration}
A cylindrical grid $\left(\rho_\mathrm{c},\phi_\mathrm{c},z_\mathrm{c}\right)$ is utilized in the simulation process, where $\rho_\mathrm{c}=\sqrt{x^2+y^2}$, $\phi_\mathrm{c}=\mathrm{atan2}(y,x)$ and $z_\mathrm{c}=z$. Indexes $i_{\mathrm{cell}}$, $j_{\mathrm{cell}}$, and $k_{\mathrm{cell}}$ are used to label grid cells. Hence, the number density of a given grid cell $(i_{\mathrm{cell}},j_{\mathrm{cell}},k_{\mathrm{cell}})$ reads
   \begin{equation}
   \begin{aligned}
      n(i_{\mathrm{cell}},j_{\mathrm{cell}},k_{\mathrm{cell}})=&\sum\limits_i\sum\limits_j\int_{r_{\mathrm{min}}}^{r_{\mathrm{max}}}p(r_\mathrm{g},i,j)\mathrm{d}r_\mathrm{g} \int_{v_{\mathrm{min}}}^{v_{\mathrm{max}}}p(v)\mathrm{d}v\\
      &\frac{n(i_{\mathrm{cell}},j_{\mathrm{cell}},k_{\mathrm{cell}},r_\mathrm{g},v,i,j)\Delta{t}}{V(i_{\mathrm{cell}},j_{\mathrm{cell}},k_{\mathrm{cell}})N_{\mathrm{start}}(r_\mathrm{g},v,i,j)}
      \end{aligned}
.\end{equation}

 For all trajectories, the grain positions are stored at equal time intervals, $\Delta t$. The symbol $n(i_{\mathrm{cell}},j_{\mathrm{cell}},k_{\mathrm{cell}},r_\mathrm{g},v,i,j)$ is the plain number of dust particles with radius $r_g$, starting velocity $v$, and starting surface element $(i,j)$ recorded in the cell $(i_{\mathrm{cell}},j_{\mathrm{cell}},k_{\mathrm{cell}})$, $V(i_{\mathrm{cell}},j_{\mathrm{cell}},k_{\mathrm{cell}})$ is the volume of the cell $(i_{\mathrm{cell}},j_{\mathrm{cell}},k_{\mathrm{cell}}),$ and $N_{\mathrm{start}}(r_\mathrm{g},v,i,j)$ is the number of grains with a given starting radius $r_\mathrm{g}$, velocity $v$, and starting element $(i,j)$.

\begin{figure}
    \centering
    \subfigure[]
    {
    \includegraphics[width=6.2cm,height=7.5cm,angle=90]{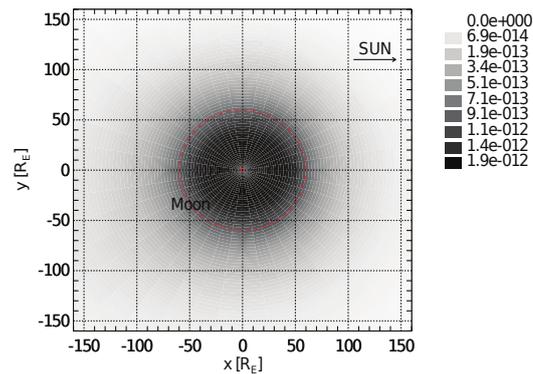}
    }
    \subfigure[]
    {
    \includegraphics[width=7.2cm,height=6.0cm]{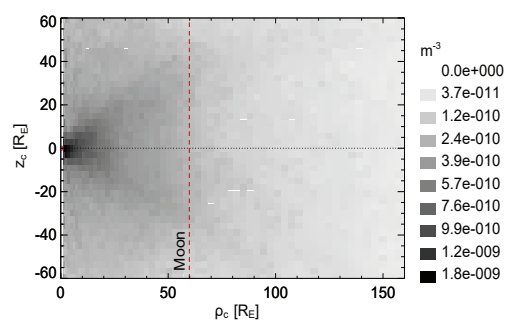}
    }
    \caption{Optical depth and number density of particles ejected from the surface of the Moon. \emph{Panel a}: Normal geometric optical depth. \emph{Panel b}: Azimuth-averaged number density in the $\rho_c-z_c$ plane. The red circle is the Earth, and the red line denotes the orbit of the Moon.}
    \label{fig: optical depth}
\end{figure}

In a similar manner we can obtain the geometric optical depth covered by the grains in a cell,
by integrating over $\pi r_g^2$ in Eq. $(21)$ and dividing by, instead of volume, the surface of
 the cell (along a given line of sight). The normal optical depth and number density distribution are illustrated in Fig.~\ref{fig: optical depth}. {Figure}~\ref{fig: optical depth}(a) shows the normal optical depth of the dust in a Sun-oriented frame, where the positive $x$ axis always points in the direction of the Sun. Particles occupy a torus between the Earth and the Moon with an outer edge that is separated by roughly one lunar Hill radius from the orbit of the Moon. The torus is asymmetric and slightly offset 
{toward} the Sun. From Fig.~\ref{fig: optical depth}(b), we can see that dust particles are distributed nearly symmetrically about the equator of the Earth, covering a wide vertical range of up to tens of Earth radii. We note, however, that the configuration shown in Fig.~\ref{fig: optical depth}(b) was evaluated in an inertial frame, averaging over grains ejected from the Moon at $12$ times equidistantly over one period of the ascending node precession of the Moon, and over an evolution time of more than $100$ years. Thus, the effects of the inclination of the Sun and the Moon are averaged out. The instantaneous dust configuration is expected to differ somewhat from the one shown in Fig.~\ref{fig: optical depth}(b) because the instantaneous Laplace plane of the dynamical problem does not coincide with the equatorial plane of the Earth.
\begin{figure}
    \centering
    \includegraphics[width=7.5cm,height=5.5cm]{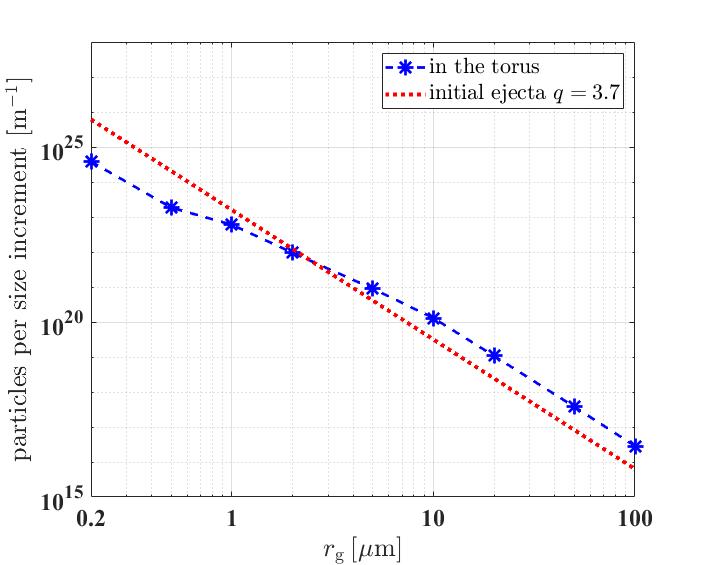}
    \caption{Steady-state differential size distribution for the torus from simulations. The red line denotes the initial ejecta size distribution from $p(r_\mathrm{g})\propto r_\mathrm{g}^{-q}$ normalized to unity, and the blue line denotes the steady-state size distribution in the torus from Fig.~\ref{fig: optical depth}.}
    \label{fig: size_distri}
\end{figure}

The differential size distribution in the torus between the Earth and the Moon is shown in Fig.~\ref{fig: size_distri}. The size distribution in the torus after long-term simulations remains close to the initial ejecta distribution but has flattened in the size range of $1$-$10 \, \mathrm{\mu m}$. This can be attributed to the rapidly changing life span for grains in this size range (see Fig.~\ref{fig: LifeSpan}(a)).

\subsection{Analysis of orbital elements}

\begin{figure}
    \centering
    \includegraphics[width=7.5cm,height=5.5cm]{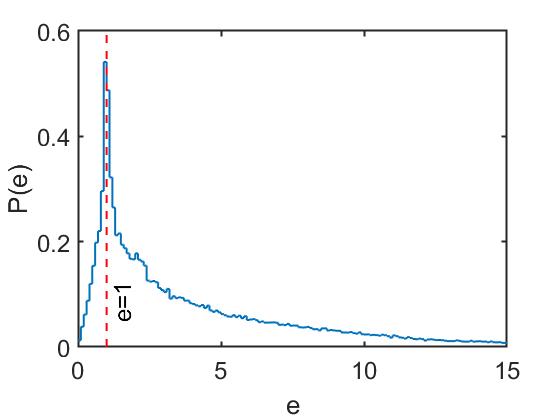}
    \caption{Distribution of eccentricity for $0.2 \, \mathrm{\mu m}$ particles.}
    \label{fig: ecc_distri_0.2}
\end{figure}
\begin{figure}
    \centering
    \subfigure[]
    {
    \includegraphics[width=7.5cm,height=5.5cm]{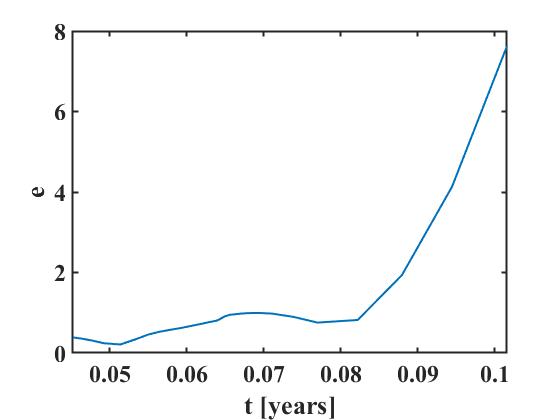}
    }
    \subfigure[]
    {
    \includegraphics[width=7.5cm,height=5.5cm]{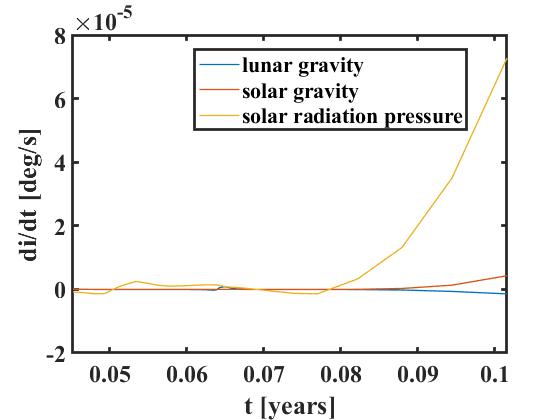}
    }
    \caption{Evolution of eccentricity and $\mathrm{d}e/\mathrm{d}t$ for one $0.2\, \mathrm{\mu m}$ particle. \emph{Panel a}: Evolution of eccentricity. \emph{Panel b}: Evolution of $\mathrm{d}e/\mathrm{d}t$.}
    \label{fig: ecc_evolution_0.2}
\end{figure}
{Figure}~\ref{fig: ecc_distri_0.2} shows the distribution of eccentricity for $0.2 \, \mathrm{\mu m}$ particles. The eccentricities are dispersed from $0$ to $15$, that is, a large number ($\sim$ $80\%$) of particles are hyperbolic, which explains their short life spans. These grains leave the Earth-Moon system in a very short time and contribute little to the number density. The evolution of eccentricity for a typical $0.2 \, \mathrm{\mu m}$ particle with a lifetime of $0.12$ years is shown in Fig.~\ref{fig: ecc_evolution_0.2}(a). The initial eccentricity is near $0.5$, {that is,} this particle starts from an elliptical orbit. Initially, the eccentricity fluctuates slightly within a few days and then grows continually, representing a typical evolution from an ellipse to a hyperbola, which is a consequence of the strong effect of solar radiation pressure. {Figure}~\ref{fig: ecc_evolution_0.2}(b) depicts the contribution of the perturbations induced by lunar gravity, solar gravity, and radiation pressure on the growth rate $\mathrm{d}e/\mathrm{d}t$, calculated from Eq.~$(5)$. Solar gravity has a negligible influence on $\mathrm{d}e/\mathrm{d}t$. Initially, the $\mathrm{d}e/\mathrm{d}t$ due to solar radiation pressure is a small negative or positive value. After 0.08 years, solar radiation pressure dominates and $\mathrm{d}e/\mathrm{d}t$ remains positive. As a result, the eccentricity increases monotonically, and finally, the orbit becomes hyperbolic.

\begin{figure}
    \centering
    \subfigure[]
   {
    \includegraphics[width=7.1cm,height=5.2cm]{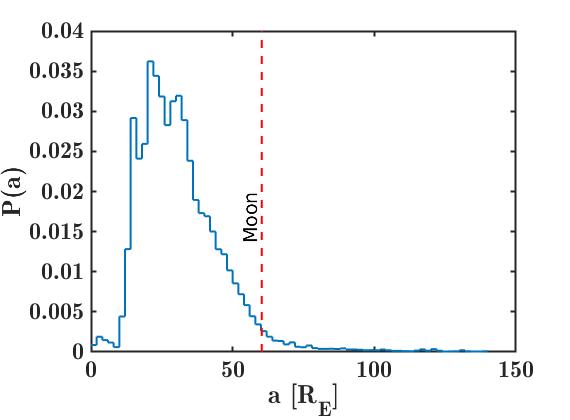}
    }
   \subfigure[]
    {
    \includegraphics[width=6.9cm,height=5.2cm]{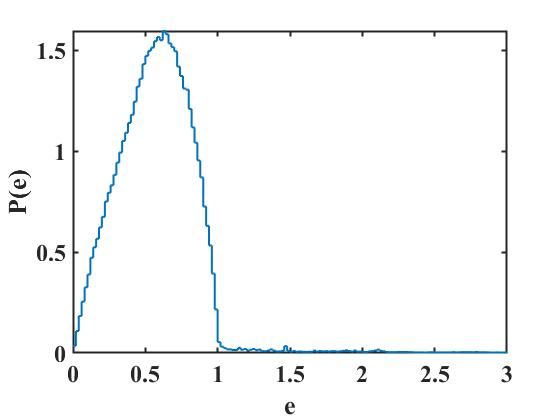}
    }
       \subfigure[]        
    {
    \includegraphics[width=7.1cm,height=5.2cm]{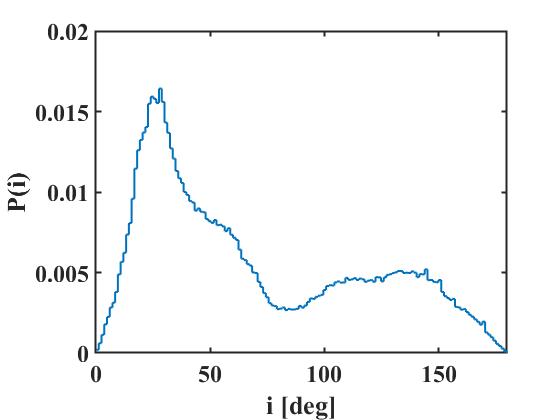}
    }
    \caption{Distribution of orbital elements for $100\, \mathrm{\mu m}$ particles ejected from the surface of the Moon. \emph{Panel a}: Distribution of the semimajor axis. \emph{Panel b}: Distribution of eccentricity. \emph{Panel c}: Distribution of inclination.}
    \label{fig: distri_100micron}
\end{figure}

Large particles remain bound to the Earth-Moon system for a much longer time, and they make a significant contribution to the dust population. The distributions of the semimajor axis, eccentricity, and inclination of all $100\, \mathrm{\mu m}$ particles are shown in Fig.~\ref{fig: distri_100micron}. The semimajor axis peaks around $20\,R_\mathrm{E}$, mainly ranging from $10\,R_\mathrm{E}$ to $50\,R_\mathrm{E}$. The eccentricity has a centered distribution below unity, peaking near 0.7. The inclination exhibits two peaks, at values of $20\degree$ and $140\degree$, {that is,} a significant fraction of grains are on retrograde orbits. 

\begin{figure}
    \centering
    \includegraphics[width=7.5cm,height=5.5cm]{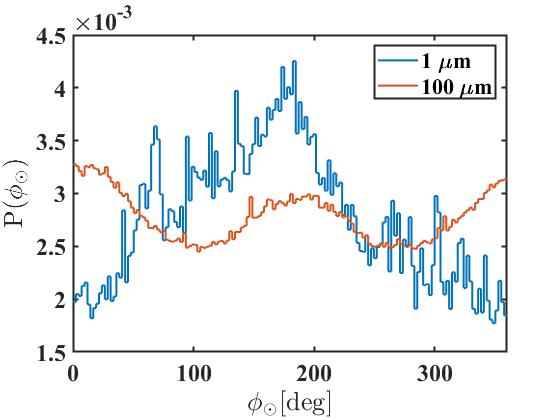}
    \caption{Distribution of solar angle for $1 \, \mathrm{\mu m}$ and $100 \, \mathrm{\mu m}$ particles ejected from the surface of the Moon.}
    \label{fig: distri_solar_angle}
\end{figure}

We also calculated the distributions of the solar angle for $1 \, \mathrm{\mu m}$ and $100 \, \mathrm{\mu m}$ grains (Fig.~\ref{fig: distri_solar_angle}). The solar angle is defined as the angle between the grain's orbital pericenter and the Sun as seen from Earth \citep{hamilton1993motion}:
\begin{equation}
\phi_\odot=\Omega+\omega-\psi_{\odot}
.\end{equation}
Here, $\Omega$ is the longitude of the ascending node, $\omega$ is the argument of periapsis, and $\psi_{\odot}$ is the solar longitude in the inertial frame.
From Fig.~\ref{fig: distri_solar_angle}, the solar angle distribution is approximately symmetric about $\phi_\odot=180\degree$. The solar angle of $1 \, \mathrm{\mu m}$ particles peaks at $180\degree$, {that is,} the population of these grains is offset toward the Sun. For $100\, \mathrm{\mu m}$ particles, the curve is smoother but shows two peaks, at $0\degree$ and $180\degree$, which indicates that the offset of the torus formed by $100\, \mathrm{\mu m}$ particles is much smaller compared to that of the $1\, \mathrm{\mu m}$ grains. For particles of different sizes, the distribution exhibits different kinds of asymmetry, with offsets toward or away from the Sun. However, averaging over grain size, the torus is slightly offset toward the Sun (see Fig.~\ref{fig: optical depth}(a)). 

\begin{figure}
    \centering
    \includegraphics[width=7.5cm,height=5.5cm]{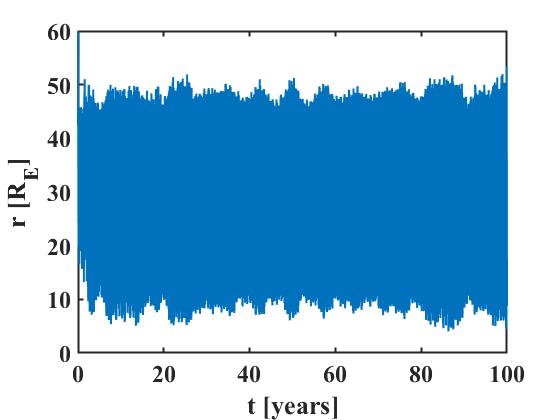}
    \caption{Evolution of $r$ for a $100\, \mathrm{\mu m}$ particle.}
    \label{fig: r_evolution_100micron}
\end{figure}

\begin{figure}
    \centering
    \includegraphics[width=7.5cm,height=5.5cm]{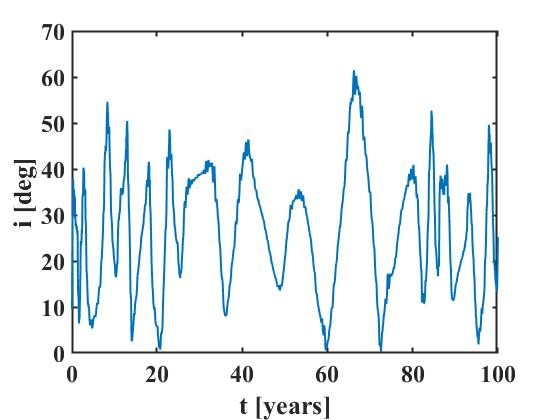}
    \caption{Evolution of inclination for a $100\, \mathrm{\mu m}$ particle.}
    \label{fig: i_evolution_100micron}
\end{figure}

Figures \ref{fig: r_evolution_100micron} and \ref{fig: i_evolution_100micron} show the evolution of radial distance and inclination for a single $100 \, \mathrm{\mu m}$ particle. This is one of the rare particles in our simulations that still remained in orbit after $100$ years. The grain covers a range between $10\,R_\mathrm{E}$ and $50\,R_\mathrm{E}$, which is consistent with the distribution in Fig.~\ref{fig: optical depth}(a). The inclination oscillates from $0\degree$ to $60\degree$. This particle is always moving on a prograde orbit within its life span. Such particles make a large contribution to the distribution of inclination (see Fig.~\ref{fig: distri_100micron}(c)).

\begin{figure}
\centering 
\includegraphics[width=7.5cm,height=5.5cm]{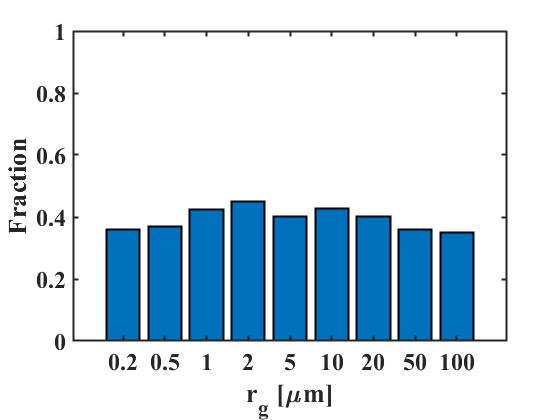}
\caption{Fraction of retrograde particles.} 
\label{fig: frac_retro} 
\end{figure}

\begin{figure}
\centering 
\includegraphics[width=7.5cm,height=5.5cm]{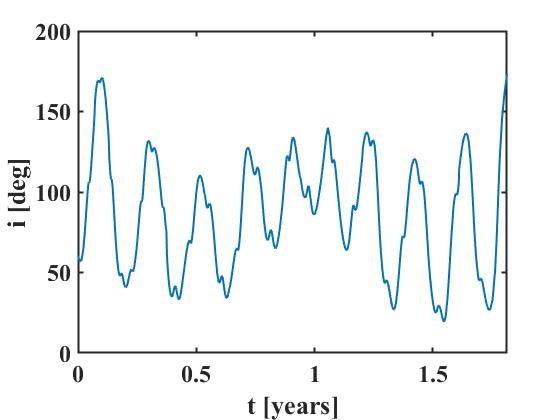} 
\caption{Evolution of inclination for a $1\, \mathrm{\mu m}$ particle.} 
\label{fig: i_evolution_1micron} 
\end{figure}

\begin{figure}
\centering
\subfigure[]
{
\includegraphics[width=7.2cm,height=5.2cm]{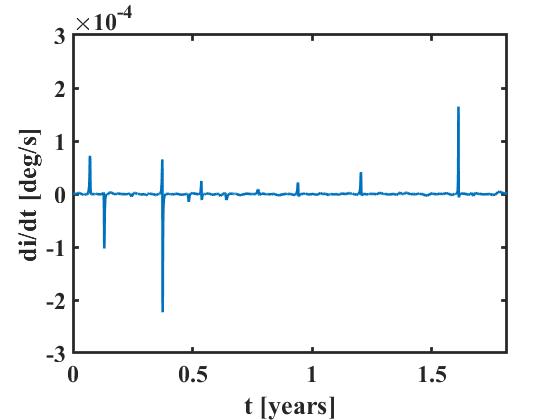}
}
\subfigure[]
{
\includegraphics[width=7.2cm,height=5.2cm]{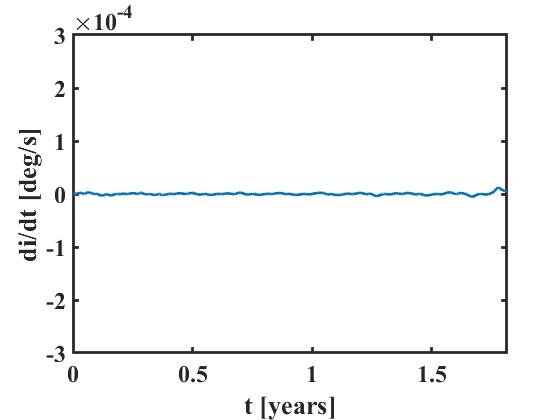}
}
\subfigure[]
{
\includegraphics[width=7.2cm,height=5.2cm]{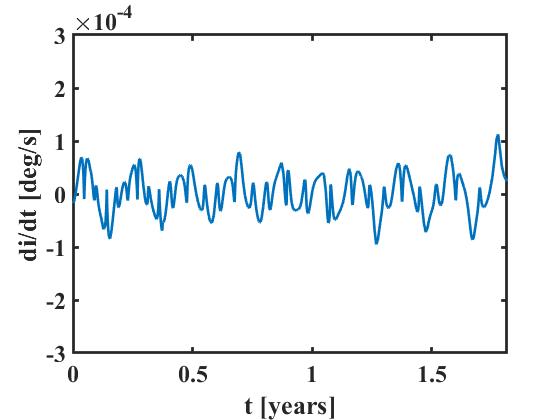}
}
 \caption{Variation in $\mathrm{d}i/\mathrm{d}t$ due to various forces for a $1 \,\mathrm{\mu m}$ particle. {\emph{Panel a}: Variation in $\mathrm{d}i/\mathrm{d}t$ due to lunar gravity. \emph{Panel b}: Variation in $\mathrm{d}i/\mathrm{d}t$ due to solar gravity. \emph{Panel c}: Variation in $\mathrm{d}i/\mathrm{d}t$ due to solar radiation pressure.}}
 \label{fig: i_variation_1micron}
\end{figure}

A considerable part of the particles in our simulations are on a retrograde orbit. The fractions of retrograde particles for different sizes are shown in Fig.~\ref{fig: frac_retro}. Some grains are ejected from the Moon directly into retrograde orbits \citep{colombo1966earth}, but a larger number of grains evolve from a prograde to a retrograde orbit in the long-term simulation. To show this, a $1 \, \mathrm{\mu m}$ particle is used here as an example. The evolution of inclination of this particle and contributions of specific forces to $\mathrm{d}i/\mathrm{d}t$ calculated from Eq.~$(6)$ are shown in Figs.~\ref{fig: i_evolution_1micron} and \ref{fig: i_variation_1micron}. From Fig.~\ref{fig: i_evolution_1micron}, we see that the initial inclination of the grain is near $60^\circ$, and the final one is $170^\circ$. This particle evolves from a prograde orbit to a retrograde orbit multiple times. As can be seen in Fig.~\ref{fig: i_variation_1micron}, the effects of gravity of the Moon and Sun on the evolution of $\mathrm{d}i/\mathrm{d}t$ are much smaller than the one induced by the solar radiation pressure, although there are some peaks from encounters with the Moon. The approximate change in inclination can be evaluated via the integration of the contribution to $\mathrm{d}i/\mathrm{d}t$ due to solar radiation pressure, which is about $100^\circ$. 

\section{Conclusions}

In this work we have derived the steady-state configuration of particles in the Earth-Moon system that have been ejected from the lunar surface in hypervelocity impacts of micrometeoroids \citep{horanyi2015permanent}. A variety of forces, including the solar radiation pressure and the gravity of the Sun, the Earth, and the Moon, are considered in a numerical exploration of the trajectories of the ejected particles in the system. Expanding on previous work in the literature \citep{Szalay2015Annual, Szalay2016Lunar, szalay2019impact}, we calculated the initial ejecta distribution and mass production rate on the lunar surface, which we used as initial conditions for the integrations. The final states (sinks), the average life spans, and the fraction of retrograde grains have been derived as functions of grain size. We also show examples for the evolution and distribution of orbital elements of particles. Small particles tend to rapidly evolve into hyperbolic orbits (on timescales of weeks), while most large particles remain in elliptical orbits for a longer time (up to a year). Many of the grains develop high inclinations, and we find that a substantial fraction evolve into retrograde orbits.

From our long-term integrations we find that about $3.6\times10^{-3}\,\mathrm{kg/s}$ ($1.8\%$) particles escape from the lunar gravity. These particles form a tenuous broad torus, the densest part of which roughly spans a distance of $40\,R_\mathrm{E}$ from the Earth. We also estimated the mass rate of particles coming into the Earth's atmosphere, $M^+_{E}=2.3\times10^{-4}\,\mathrm{kg/s}$, which is a non-negligible value. As the space activities in the Earth-Moon system are much more frequent compared with other planets, dust around the Earth seriously threatens not only the mechanical structure of spacecraft but potentially also the health of astronauts. In Apollo missions, astronauts were exposed to the lunar dust environment, and respiratory, dermal, and ocular irritations from lunar dust were reported \citep{turci2015free}. NASA established the Lunar Airborne Dust Toxicity Advisory Group (LADTAG) to 
assess the risk degree for spacecraft and astronauts in lunar missions \citep{khan2008lunar}. Thus, analysis of the characters for particles in the torus around the Earth is of great significance for assessing the space environment and ensuring the security of explorations. 

We characterize the torus in terms of the normal geometric optical depth and number density. Owing to the effect of solar radiation pressure, the torus is mildly offset toward the solar direction. The peak normal optical depth we obtain is on the order of $10^{-12}$, which is small compared to the optical depth of the Thebe extension of Jupiter's gossamer rings, $\sim10^{-9}$ \citep{2018prs..book..125D}, and the limits inferred for the yet undetected Phobos ring of Mars, $\sim10^{-8}$ \citep{2006P&SS...54..844S}. The peak number density we find for the lunar torus is on the order of $10^{-9} \,\mathrm{m}^{-3}$, which is about four orders of magnitude smaller than the density of the Deimos ring inferred for radii of $15\,\mathrm{\mu m}$ by \citet{juhasz1995dust} and four and five orders of magnitude smaller than the number for the Deimos and Phobos rings ($\geq 0.5\,\mathrm{\mu m}$) determined by \citet{liu2021configuration}, respectively. For an in situ detector on a circular orbit at a distance of $10\,R_\mathrm{E}$ pointing in the prograde orbit direction, this translates into a flux of about $50$ grains per $\mathrm{m}^2$ and per year (taking into account that a fraction of the grains are on retrograde orbits). From the Interplanetary Meteoroid Engineering Model \citep[IMEM;][]{dikarev2005new}. we find that this is a factor of several smaller than the flux of micron-sized interplanetary micrometeoroids at $1\mathrm{AU}$. The directional distributions of the lunar grains and the interplanetary particles, however, would be different.

Our results on the grain lifetimes, their sinks, and their typical orbital evolution are robust because the relevant perturbation forces are taken into account and the respective parameters (masses of the gravitational perturbers, the solar constant, and the solar radiation pressure efficiency factor) are known up to small uncertainties. Large uncertainties, however, remain in our calibration of number density and optical depth. On one hand, our simulation results are based on the mass production rate of $M^+_\mathrm{total}=0.2\,\mathrm{kg/s}$ reported by \citet{szalay2019impact}. If we follow the method used in the literature for an analysis of ejecta from the Galilean moons \citep{krivov2003impact, 2003P&SS...51..455S}, employ the mass flux at 1AU from the IMEM model \citep{dikarev2005new}, and use the empirical expression for the yield suggested by \citet{koschny2001impactsa}, we obtain a mass production rate that is about one order of magnitude higher. Although in our analysis we prefer to use the value from the direct measurement, this suggests that the uncertainty in the calibration of our model is at least one order of magnitude. On the other hand, we assume a power law distribution for the ejection velocity, whose high velocity tail determines the fraction of ejecta that escape the Moon in our model. In the literature, a different functional form was fitted successfully to the bound particles detected by LDEX \citep{Szalay2016Lunar}. If the tail of the velocity distribution of the lunar ejecta turns out to be different from the one assumed in our paper, then the dust densities derived from our model will also change.

\begin{acknowledgements}
      This work was supported by National Natural Science Foundation of China (No. 12002397) and by the China Scholarship Council (CSC, 201906220134). We acknowledge CSC -- IT Center for Science for the allocation of computational resources. 
\end{acknowledgements}

%
%

\newpage
\bibliography{ref.bib}
\bibliographystyle{aa}

\end{document}